%% file: manuscript.tex
\documentclass{article}

\usepackage{arxiv}
\usepackage{listings, color}
\usepackage[square,sort,comma,numbers]{natbib}
\usepackage{authblk}
\usepackage[table,dvipsnames]{xcolor}
\usepackage[utf8]{inputenc} 
\usepackage[T1]{fontenc}    
\usepackage{hyperref}       
\usepackage{url}            
\usepackage{booktabs}       
\usepackage{amsfonts}       
\usepackage{nicefrac}       
\usepackage{microtype}      
\usepackage{cleveref}       
\usepackage{lipsum}         
\usepackage{graphicx}
\usepackage{natbib}
\usepackage{doi}
\usepackage{multirow}
\usepackage{tikz}
\usepackage{hyperref}
\usepackage{multirow}
\newcommand{\dslkeyword}[1]{\texttt{\small #1}}

\title{A DSL for Defining Feature-Level Quality Constraints and the Aggregation of Evaluation Results in DevOps}

\lstdefinestyle{atg}{%
     basicstyle=\ttfamily,       
     frame=single,               
     framesep=1pt,               
     framerule=1pt,            
     breaklines=true,            
     breakindent=0pt,             
     backgroundcolor=\color{gray!10}
}

\date{}

\author[1,2]{Philipp Haindl}
\author[2]{Reinhold Plösch}
\affil[1]{Software Competence Center Hagenberg, Austria}
\affil[2]{Department of Business Informatics - Software 
Engineering, Johannes Kepler University Linz, Austria}

\renewcommand{\headeright}{}
\renewcommand{\undertitle}{Technical Report}
\renewcommand{\shorttitle}{A DSL for Defining Feature-Level Quality Constraints and the Aggregation of Evaluation Results in DevOps}

\hypersetup{
pdftitle={A template for the arxiv style},
pdfsubject={q-bio.NC, q-bio.QM},
pdfauthor={Philipp Haindl, Reinhold Plösch}
}

\begin{document}

\maketitle

\begin{abstract}
Quality requirements typically differ among software features, e.g., due to different usage contexts of the features, different impacts of related quality deficiencies onto overall user satisfaction, or long-term plans of the developing organization. For instance, maintainability requirements might be particularly high for software features which are frequently used or bear strategic value for the developing organization. Also, software features where even the smallest delays are perceived as negative by the user will be subjected to specially tight performance requirements. 

We defined an operational DSL to define software quality requirements as individual feature-level constraints based on quantitative measures. The DSL provides language elements to define the operationalization of measures from external systems, time series operations, time filters, and the automatic evaluation of these feature-level constraints in DevOps based on comparison operators and threshold values. In addition, quality ratings summarize evaluation results of features on an ordinal grading scheme. Likewise, quality gates use these quality ratings to reflect the fitness of software features or the overall software product using different states. Finally, we show an example based on a widely-adopted secure mobile messaging app that illustrates the interplay of the different DSL elements.
\end{abstract}


\urlstyle{rm}

\section{Introduction}
\input{introduction}

\section{Related Work}
\input{related_work}

\section{Concepts of the TAICOS Approach}
\input{concepts}

\section{Specifying Feature-Level Quality Constraints}
\input{constraint_language}

\section{Defining the Aggregation of Constraint Evaluation Results}
\input{aggregation_language}

\section{Illustrative Example}
\input{example}


\section{Discussion}
\input{discussion}

\section{Conclusion and Future Work}
\input{conclusion}

\bibliographystyle{IEEEtran}
\bibliography{library}

\end{document}

%% file: introduction.tex
\label{introduction}
In most software projects, quality requirements are approached as cross-cutting concerns on the top-level of the software product and treated without further refinement for individual software features. This, however, neglects that the same quality requirement might manifest differently among individual software features. As an example, maintainability requirements might be higher for those software features that need to be changed more frequently than others, e.g., because they represent core functionality of the software product or differentiate the product from competitors' alternatives.  Likewise, software features that process real-time data will be subjected to stricter performance requirements than features that allow to administer customer master data. To permit the specification of individual quality requirements for different software features and the evaluation of these feature-level quality requirements in DevOps, we developed the \textit{TAICOS approach}. In \cite{haindl_extension_2019} we presented a software quality model that allows to specify feature-level quality requirements and introduced the concept of \textit{feature constraints} to express the fulfillment criteria for these quality requirements. This paper is an extension of our previous presentation \cite{haindl_operational_2020} of an operational DSL to specify these feature constraints and the software prototype to operationalize the required measures  to evaluate the fulfillment of these feature constraints in DevOps. 

An operational DSL that allows automated quality assessments in DevOps must permit to specify the evaluation criteria of feature constraints, the operationalization of measures used in the constraints, and the suitable aggregation of evaluation results in an operational manner, individually for different features. In addition, such a DSL needs to address two major challenges: First, it must regard the different types of measures that typically accrue on the systems involved in DevOps and which are necessary to evaluate the respective constraints. Second, as these measures evolve over time and change frequently, an appropriate DSL also needs to provide capabilities to express the permissible development of a measure in a particular time frame. 

From the viewpoint of the operationalization of such a DSL, the lead time for evaluating feature constraints must be minimal even for large-dimensioned features as well as for constraints taking into account a multitude of measures. This specially affects the application of the DSL in DevOps, which demands fast and frequent feedback cycles covering large data streams originating from multiple different development and operational systems. The software prototype presented in \cite{haindl_operational_2020} addresses these particularities pertaining to the acquisition, preprocessing, filtering, and calculation of measures required for constraint evaluation. Also, the validation of this software prototype through a lab experiment showed that it provides the evaluation results of feature constraints of different scopes in a timely fashion that is suitable for the DevOps context.

In general, DSLs provide language elements for expressing the requirements of a specific problem domain, using the concepts suitable for that particular domain \cite{le_domain_2018}. In the context of the presented operational DSL this means that software quality requirements can be expressed on feature-level by domain experts. In \cite{haindl_specifying_2020} we reported on a case study with practitioners to assess the completeness, expressiveness, scope, and suitability of the DSL for specifying maintainability requirements on feature-level as one important software quality aspect. While the participants rated the aforementioned aspects of the DSL overall with high levels, the experts also stressed the necessity of expressing the aggregation of  constraint evaluation results.

To address this challenge, we extended the DSL by two novel elements that allow to specify the aggregation of constraint evaluation results through \textit{quality ratings} and \textit{quality gates}. As an important extension of our previous paper \cite{haindl_operational_2020} in this context, this paper elaborates the different language elements of the DSL in much greater detail and also introduces the two new additions for aggregating constraint evaluation results. Also, it discusses our findings from two former case studies focusing on different aspects of this DSL and the respective software prototype for its practical integration in DevOps, from a broader perspective. Finally, in this paper we also show visualizations of the new concept for aggregation of constraint evaluation results through our software prototype with a running example.

The paper is structured as follows: In Section \ref{related_work} we elaborate on the related work in this field and the demarcation to our approach. Section \ref{concepts} describes the different concepts of the TAICOS methodology for getting the bigger picture of our research. In Section \ref{constraint_language} we present the different language elements of the DSL for specifying feature-level quality constraints, followed by the presentation of language elements for defining the aggregation of evaluation results on different levels in Section \ref{aggregation_language}. Subsequently, in Section \ref{example} we give an illustrative example of the new results aggregation concept from our software prototype. Following, in Section \ref{discussion} we discuss the presented DSL and particularly the two new language elements for results aggregation before the background of the previous case studies. Finally, we sketch possible directions for future work and conclude our paper in Section \ref{conclusion}.

%% file: related_work.tex
\label{related_work}
We separate the related work in this field into three streams of research. The first stream relates to constraint languages for expressing and evaluating the fulfillment of quality requirements. Event-based constraint languages \cite{robinson_monitoring_2002,robinson_roadmap_2010,robinson_requirements_2006} mainly target on monitoring and assuring special runtime requirements of systems, e.g., the proper message exchange and failover in systems-of-systems (SoS). They rely on monitoring \textit{discrete} events and thus are less suitable for monitoring the fulfillment of \textit{continuous} quality requirements, such as e.g., maintainability requirements. Vierhauser et al. describe a DSL-based approach for event monitoring in SoS \cite{vierhauser_developing_2015}. The language allows defining event triggers, conditions, and timing requirements for these conditions. Rosenberg et al. present the \textit{Vienna Composition Language (VCL)} \cite{rosenberg_end--end_2009}, a DSL for QoS-aware web service composition. It permits to specify QoS attributes e.g., response time or availability, for individual services. Lastly, Goknil and Peraldi-Frati developed a DSL that provides dedicated language elements for expressing time spans, jitter, minimum, and maximum of measures \cite{goknil_dsl_2012}.

The second stream of related research addresses the aggregation of software quality evaluation results. Siavvas et al. present the \textit{QUATCH} framework for software product quality assessment \cite{siavvas_qatch_2017}. In the approach, multiple metric-based quality assessments are summarized in a single score, which is subsequently mapped onto a rating. While the score is a metrical value that lies in the interval between 0 and 1 (with 0 reflecting lower quality), the rating is an ordinal representation of this metrical value. The relation between score and rating is linear, i.e., the higher the score, the higher the rating on its ordinal scale. To reflect the individual importance of the metrics as the basis of the quality score, each metric can be assigned a different weight. The quality score is then calculated from the weighted average of these metrics. Apel et al. elaborate a metric-based quality rating approach for microservice architectures \cite{apel_towards_2019}. In contrast to Siavvas et al., these authors do not calculate a quality rating per se, but examine the suitability of quality metrics relevant for microservice architectures to rate the quality of each of the eight ISO 25010 subcharacteristics. This also follows the idea of summarizing multiple individual metrics into one single quality score, e.g., through building the average of these metrics. A quality rating approach that takes into account threshold values for individual metrics is presented by \cite{bansal_software_2016}. Thereby, to fulfill a certain rating (similar to school grades) individual metrics need to lie within a specific predefined interval. 

The third stream of related research pertains to the application of quality gates in software projects. The idea of quality gates goes back to the ``stage-gate process'' first presented by Cooper \cite{cooper_perspective_1994,cooper_product_1999} to assert the product quality at important transitions between product development phases. At these transition, \textit{stage gates} assess whether the preceding product phase meets particular criteria so that the development can proceed to the next stage. Besides other areas, the idea is applicable also to agile software product development \cite{borba_agile_2019,bianchi_agile_2020} to ensure software quality between sprints. Raatikainen et al. report on an exploratory multiple case study to examine different stages, controlled by gates, in lean software development in startups \cite{raatikainen_eight_2016}. They propose four different stages and respective gates, from ideation to scaling, to ensure the economic viability of the developed software product. A further, industrial case study investigated the application of the stage gate concept to agile software development \cite{bianchi_agile_2020}. The authors stress the importance of defining stage gates that take into account metrics from different domains and also the requirement to flexibly adapt these gates, particularly in an agile context, throughout an agile project. Thus, stage gates shall be composed of fine-grained metrics typical for the preceding development phase. 

In the remainder of this paper we refer to ``stage gates'' as quality gates, as also the further cited literature uses this term. Flohr examines the appropriateness of criteria used in quality gates to help decide whether quality milestones in projects are fulfilled \cite{flohr_defining_2008}. He concludes that systematic-down approaches based on software quality models and deriving criteria from abstract business goals are must suitable for the definition of quality gates. However, as these findings originate from a validation through student experiments this might affect the generalizability of these findings. Schermann et al.  report on a case study with companies to examine the benefits of using quality gates for continuous delivery of software projects \cite{schermann_towards_2016}. The case study revealed a relation between the effort companies put into quality gates and the pace with which they can release new versions of their software. In a subsequent work of these authors \cite{schermann_were_2018}, this observation is explained with the intrinsic goal behind continuous delivery to increase velocity. This pertains to the time needed to pass all quality gates and approval steps until the changes to the software reaches the production environment, while at the same time ensuring a high level of software product quality. Deniz gives a concrete example of quality gates that focus on object-oriented design quality based on the metrics suite of Chidamber \& Kemerer \cite{chidamber_metrics_1994} for object-oriented design \cite{deniz_software_2015}. It determines one of three states of quality gates (success, warning, fail) based on the number of quality-related rule violations. Each metrics must lie within specific thresholds to reach one of the three states. The overall quality gate of software components is determined by assigning points (100 for success, 50 for warning, 0 for fail), building the average of these points, and determining one of the three states thereof.

\subsection{Summarization of Limitations}
The main limitations of existing DSLs for evaluating quality requirements can be summarized as follows: First, they do not allow to define quality requirements individually for different software features. Second, they are mainly suitable for defining quality requirements that are based on discrete events. Third, other DSLs do not explicitly address the operationalization of the measures required for evaluating the fulfillment of the described requirements. Fourth, they provide only basic support for time-filtering and time-series analysis of constraint measures.

Pertaining to the aggregation of quality evaluation results in general, it can be observed that different approaches exist for defining quality ratings and gates. However, none of these approaches is based on a declarative language and instead are implemented in a project-specific manner. As such it is not possible to adapt the calculation of quality ratings and gates, e.g., to take into account other quality metrics or customize the weight of metrics. Adjusting the weight, for instance, is necessary to reflect the different importance of a metric in the quality score, which is calculated through the weighted average. Further, as also pointed out in  \cite{siavvas_qatch_2017}, the recent advent of machine learning approaches for software quality rating yet allows to calculate quality ratings, but in a way that is mostly intransparent and difficult to customize in practice. These approaches also do not allow to incorporate domain-knowledge available in the companies to formulate quality gates and ratings.

\subsection{Contributions}
In this paper, we contribute an operational DSL language that addresses the aforementioned limitations. Particularly, as a demarcation to related works in this field, our DSL  allows to
\begin{itemize}
	\item specify quality requirements of software features using quantitative measures and evaluation criteria as feature constraints
	\item define the operationalization of these measures, i.e., their acquisition from the various data sources in DevOps during constraint evaluation
	\item filter measures used in a feature constraint by time frame
	\item analyze time series of measures
	\item specify quality ratings flexibly in different granularity levels based on constraint evaluation results
	\item define quality gates on software feature- and product-level based on fulfillment criteria of quality ratings on different granularity levels
\end{itemize}

%% file: concepts.tex
\begin{figure*}[h]
\centering
\includegraphics[scale=0.94]{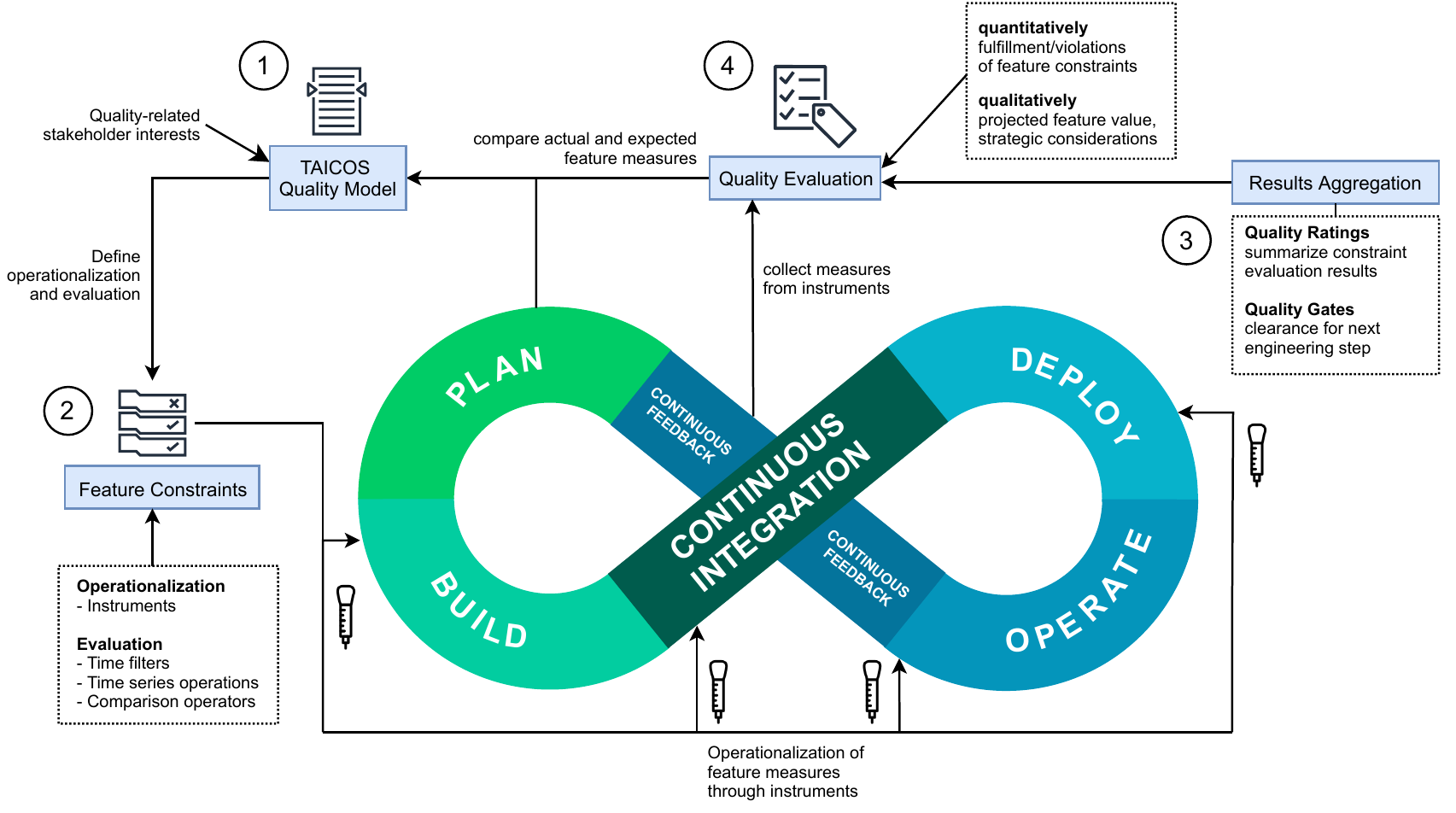}\caption{Integration of the TAICOS approach in DevOps.\label{fig:taicos_concepts}}
\end{figure*}

\label{concepts}
For gaining a better understanding of the TAICOS approach, we give a brief overview of the most important concepts shown in Figure \ref{fig:taicos_concepts}. Our approach introduces the concept of \textit{stakeholder interests} to denote top-level non-functional objectives articulated by stakeholder that must be taken into account by all software features. In general, these stakeholder interests are not limited to technical quality-related objectives, but also comprise business-related objectives of stakeholders towards a software system. In the context of this paper, we concentrate on quality-related stakeholder interests solely. The starting point for applying the approach in a DevOps environment is the (1) TAICOS quality model \cite{haindl_extension_2019} on the upper left part of the figure. This quality model allows to specify the stakeholders' interests as quality requirements on the level of individual software features and is created typically during the \textit{plan} phase of the DevOps cycle. 
The feature-level quality requirements are decomposed and defined using the TAICOS DSL as (2) feature constraints. Constraints are composed of different quantitative measures which are operationalized through instruments. During the \textit{build}, \textit{continuous integration}, \textit{deployment}, and \textit{operation} phase of the DevOps cycle, these measures are acquired from the different systems automatically from software connectors associated with these instruments. For specifying the evaluation of these measures, the TAICOS DSL provides different time filters, time series operations, and comparison operators. In the context of the \textit{continuous feedback} phase of the DevOps cycle, the measures are collected from the different instruments and the fulfillment of the feature constraints evaluated based on these measures. The individual constraint evaluation results are (3) aggregated to facilitate their interpretation in the context of a particular software feature or the overall software product. Thereby, quality ratings summarize multiple constraint evaluation results based on a five-element grading scheme. Quality gates can take three states to reflect the appropriateness of a feature or the overall software product for the next engineering step, e.g., productive deployment, system testing, or feature experimentation \cite{kevic_characterizing_2017,olsson_hypex_2014}, by taking account these quality ratings. The final (4) quality evaluation has two directions: The quantitative evaluation is done automatically through the TAICOS software prototype and takes into account constraint violations and the thereof calculated quality ratings and the status of quality gates. The qualitative evaluation needs to be done manually among the relevant stakeholders of the software project and takes into account the projected feature value for the developing organization and the end user of the software. This evaluation is based on strategic and economic considerations impacting the \textit{software value chain} of a company \cite{wirtz_business_2011,pussep_software_2012} and thus cannot be conducted in an automated fashion. In case of important findings from any of these two directions of evaluation, the TAICOS quality model is adapted accordingly. For instance, by relaxing or enforcing existing feature constraints or by adding new ones to better suit the overall quality-related stakeholder interests.

%% file: constraint_language.tex
\label{constraint_language}
As already stated, the TAICOS approach allows to decompose quality-related stakeholder interests into feature-level quality requirements and specify them in an operational manner as feature constraints. Thereby, each feature constraint utilizes a specific \textit{measure} that is operationalized through an \textit{instrument} and accessed in the constraint declaration by a \textit{variable}. As an example, we could utilize the cyclomatic complexity as a measure of the maintainability of the software. This measure can be directly acquired, e.g., from the \dslkeyword{Sonar} instrument\footnote{https://www.sonarqube.org/} which queries this measure from the corresponding code quality server and assign the value to a variable. The concrete feature-level requirement towards the maintainability and thus, the permissible feature-dependent threshold for this variable, is then expressed as a feature constraint. To evaluate the development of a measure over time, the TAICOS DSL also provides time filters and time series operations. For instance, in the given example one can demand that the average cyclomatic complexity during the last month must be below a certain threshold. Conceptually, variables can hold metrical values (e.g., the number of rule violations or concrete metrics), or ordinal values such as quantiles resulting from benchmarking rule violations with other projects.

\subsection{Grammar and Language Elements}
\label{warningThreshold}
In general, each TAICOS model consists of a project-specific metadata block and the four blocks for specifying measurements and constraints: \dslkeyword{<Interests>}, \dslkeyword{<Features>}, \dslkeyword{<Variables>}, and a list of \dslkeyword{<Constraint>}. The metadata block specifies the unique project identifier (\dslkeyword{projectId}) and the  \dslkeyword{<WarningThreshold>}. This threshold denotes the limit in percent below which the user shall be warned in advance, e.g., on a user interface, of possible constraint violations. Listing \ref{lst:metadata} shows the grammar rules for the model metadata block using the DSL. 
\begin{figure}[h!]
\lstset{style=atg,basicstyle=\footnotesize\ttfamily,caption={Grammar rules for defining TAICOS model metadata.},captionpos=b,label=lst:metadata}
\begin{lstlisting}
"projectId: " projectId "." .
<WarningThreshold> ::= "at warning" warningThreshold 
    "before constraint violation." . 
\end{lstlisting}
\end{figure}

In listings of grammar rules, we highlight \dslkeyword{<non-terminal symbols>} of the language in angle brackets and \dslkeyword{terminal symbols} without brackets. 

\subsubsection{Variables and Instruments}
\label{variabledefinition}
\label{instrumentdefinition}
A \dslkeyword{<Variable>} defines a name for a concrete measure, its type, and also its acquisition from an \dslkeyword{<Instrument>}. The name of a variable can be arbitrarily chosen and can be used in arbitrary constraints. The language allows three types of variables: \dslkeyword{measure} for numerical metrics, \dslkeyword{rule} for the number of rule violations, and \dslkeyword{rating} for 1-letter quality ratings on an ordinal scale from A to E. The type of a variable also imposes restrictions on the use of the \dslkeyword{comparison} operator, as only variables of the same type can be compared. 

\begin{figure}[h!]
\lstset{style=atg,basicstyle=\footnotesize\ttfamily,caption={Grammar rules for defining variables and instruments.},captionpos=b,label=lst:variablesinstrument}
\begin{lstlisting}
<Variables> ::= "@variables: {"  
    <Variable> { <Variable> } "}." .
<Variable> ::= variable 
    "as" type "from" <Instrument> "." .
type := "measure" | "rule" | "rating" .
<Instrument> ::= instrumentId  
    "(" key ["," param] ")." .
\end{lstlisting}
\end{figure}

An \dslkeyword{Instrument} defines for each variable how it can actually be retrieved from a specific external system. Also, the set of instruments can easily be extended by implementing a specific software interface to integrate further external system for retrieving measures. One could, e.g., declare a variable  \dslkeyword{cyclomaticComplexity as measure from Sonar("complexity")}. This expresses that the metrical variable \linebreak \dslkeyword{cyclomaticComplexity} is acquired from the \dslkeyword{Sonar} instrument, which accesses it using the key \dslkeyword{complexity}. Thereby, the \dslkeyword{key} used by the instrument is predetermined by the external system that the instrument connects to.

\subsubsection{Features} 
\label{featuredefinition}
The \dslkeyword{<Feature>} element specifies a software feature through a unique \dslkeyword{featureId} and a list of contexts each with an individual \dslkeyword{contextId}. In DevOps, quality-related data need to be acquired from different systems, e.g., code repositories, bug databases, application monitoring systems, etc. As features are represented differently by these systems one needs to specify how to identify a feature individually for each system (context). Context-specific metadata are defined for each feature through \dslkeyword{key}-\dslkeyword{value} pairs (see Listing \ref{lst:features}).

\begin{figure}[h!]
\lstset{style=atg,basicstyle=\footnotesize\ttfamily,caption={Grammar rules for defining feature metadata.},captionpos=b,label=lst:features}
\begin{lstlisting}
<Features> ::= "@features: {" 
    <Feature> { <Feature> } "}." .
<Feature> ::= featureId ":" 
    ["with quality-rating" ratingId 
        ["and quality-gate" gateId]
    ]
    ["with quality-gate" gateId ]
    {"context" contextId "with {" 
            { key value "." } 
        "}" 
    }"." .
\end{lstlisting}
\end{figure}
As an example, one could declare a context \textit{filesystem} to denote the source code location of the software feature on the developer's computer, a \textit{git} context to define the repository location of  the versioning system, or an \textit{apm} context to query runtime measures from a system for application performance monitoring. Due to the genericity of this concept, different contexts and thus sources for retrieving feature-dependent quality metrics can be declared. 

\subsubsection{Constraints}
\label{typeofconstraints}
\begin{figure}
\lstset{style=atg,basicstyle=\footnotesize\ttfamily,caption={Grammar rules for defining feature constraints.},captionpos=b,label=lst:featureconstraints}
\begin{lstlisting}
<Constraint> ::= interestId ": {" 
    featureId ": {" <ConstraintBody> 
    {"," <ConstraintBody> } ["as" constraintName] "}." .
<ConstraintBody> ::= 
    ( variable [".benchmark"] ) | 
    ( operation "(" variable "," filter ")" )
    comparison constraintThreshold "." .
operation ::= "min" | "max" | "median" | "avg" | "gradient".
filter ::= "seconds" | "minutes" | "hours" | "days" | "months".
comparison ::= ">" | "<" | "==" | ">=" | "<=".
\end{lstlisting}
\end{figure}
A \dslkeyword{<Constraint>} specifies a quantitative requirement for a criterion, e.g., cyclomatic complexity. The evaluation criteria are expressed in the \dslkeyword{<ConstraintBody>}. Optionally, a constraint can be given a unique \dslkeyword{constraintName} to reference it in definitions of quality ratings (cf. Section \ref{aggregation_language}).
Listing \ref{lst:featureconstraints} shows the grammar rules for feature constraints using time series operations, time filters, and comparison operators. Basically, one can differentiate three variants of constraints: 
\begin{enumerate}
	\item Constraints comparing the most recent value of a variable with a threshold. This is the simplest form of a constraint and works with any type of variable. For instance, one could declare a constraint such as \dslkeyword{video: cyclomaticComplexity < 8} to require that cyclomatic complexity for the software feature \dslkeyword{video} is lower than 8. In this constraint variant, the \dslkeyword{threshold} element can  take metrical values, e.g., the expected cyclomatic complexity or the permissible number of rule violations, or ordinal values expressed as single characters from A to E. 

	\item Constraints utilizing benchmark results of rule violations to evaluate if they lie within a specific quartile of the benchmark base. This type of constraint requires a variable of type \dslkeyword{rule} and is declared by appending \dslkeyword{.benchmark} to the variable name. One could, e.g., declare a variable \dslkeyword{undocumentedMethods as rule from Sonar("squid:UndocumentedApi")} to hold the number of violations of the rule to document public methods. The constraint \dslkeyword{video: undocumentedMethods.benchmark < Q50} would then require that the number of rule violations in the \dslkeyword{video} feature is lower than in 50\% of the benchmarked projects. Thereby, the \dslkeyword{threshold} can take any of the four quartiles which are referred to as \dslkeyword{Q25}, \dslkeyword{Q50}, \dslkeyword{Q75}, and \dslkeyword{Q100}. 

	\item Constraints comparing the result of a time series operation, applied onto measures, with a concrete threshold.  Such constraints can only be expressed for metrical variables, i.e., of type \dslkeyword{measure} or \dslkeyword{rule}. They require a concrete time series \dslkeyword{operation} and a time \dslkeyword{filter} to be specified in the constraint. The constraint \dslkeyword{video: avg(cyclomaticComplexity, days(7)) < 12} defines that the average cyclomatic complexity of the \dslkeyword{video} feature must not have been greater than or equal 12 during the last 7 days. Thereby, the \dslkeyword{threshold} can only take metrical values as the mathematical operations underlying the time series operations are only defined for real numbers. 
\end{enumerate}

In general, threshold values are always given without unit of measurement as this would require conversion between the units of variable and threshold values. Thus, the unit of measurement used for comparison is determined by the instrument that operationalizes the corresponding variable.

\subsubsection{Interests}
Value expectations that shall be regarded by all software features are specified through an \dslkeyword{<Interest>} element with a unique \dslkeyword{interestId} and a textual \dslkeyword{interestDescription} of the objective (see Listing \ref{lst:interests}).
\begin{figure}[h!]
\lstset{style=atg,basicstyle=\footnotesize\ttfamily,caption={Grammar rules for defining stakeholder interests.},captionpos=b,label=lst:interests}
\begin{lstlisting}
<Interests> ::= "@interests: {" 
    <Interest> { <Interest> } "}." .
<Interest> :: = interestId 
    ["with quality-rating" ratingId] ":" 
    interestDescription ".".
\end{lstlisting}
\end{figure}
 Interests can be nested into each other, for example to refine one larger interest into several smaller ones. In that case, the overarching large interest is referred to as \textit{interest group}. For each interest group and interest, a quality rating can be defined that summarizes the evaluation results of the constraints specified for this interest.
\subsubsection{Time Series Operations and Filters}
The DSL provides five different time series operations, like the \dslkeyword{min} and \dslkeyword{max} operation to determine minimum and maximum, the \dslkeyword{median} and \dslkeyword{avg} for calculating median and average, and the \dslkeyword{gradient} operation to calculate the incline of the (linear regression) line of a time series of measures. For filtering time series the language provides five filters such as \dslkeyword{seconds}, \dslkeyword{minutes}, \dslkeyword{hours}, \dslkeyword{days}, and \dslkeyword{months} with a numerical parameter indicating the number of time units to go backwards in time. Time series operations are only applicable for variables with a type of \dslkeyword{measure} or \dslkeyword{rule}. Also, the comparison of a variable with a threshold inevitably requires that both values are of the same type.

\subsection{Comparability of Measures in Feature Constraints}
While the syntatic validity of feature constraints formulated by the user is assured by the parser of the  DSL, it cannot check whether the measures used in the constraints are semantically appropriate to be compared. This particularly pertains to the
\begin{itemize}
\item required normalization of quality-related measures to regard the different code bases of features, and 
\item direction of comparison operators and threshold values. 
\end{itemize}
To address the first point, the TAICOS DSL allows to provide optional parameters in the declaration of an \dslkeyword{instrument}. Information required by an instrument for normalization can that thus be passed by a specially named parameter, whereby this name can freely be determined by the instrument. For instance, the \dslkeyword{Sonar} instrument used in the previous example can require a parameter \dslkeyword{normalizationBase=loc} to determine that measures of static code analysis shall be normalized based on lines of code or \dslkeyword{normalizationBase=methodCount} to normalize them based on method count.

The second point relates to the semantically appropriate selection of comparison operators. The choice of the semantically correct comparison operator depends on whether the value of the compared metric should be low or high in order to have a positive impact on quality. For instance, while the value of particular metrics shall always be as \textit{low} as possible (e.g., the technical debt) it is nonetheless possible to demand that this value is \textit{higher} than a given threshold and apply the \dslkeyword{>} operator. As the TAICOS DSL parser does not detect such semantic contradictions it is up to the user to select an operator that compares values and metrics from the background of positive impact on the quality objective, and thus stakeholder interest.

%% file: aggregation_language.tex
\label{aggregation_language}
In the frame of the development of the TAICOS DSL we already conducted a case study with practitioners \cite{haindl_specifying_2020} who applied the DSL to formulate feature-level maintainability constraints. One important finding of this case study was the practical need articulated by participants of a formalism to aggregate constraint evaluation results. To address this practical requirement, we extended the TAICOS DSL to support the definition of quality ratings and gates. Thereby, the underlying calculation approach provides for default quality ratings, which are calculated even if they have not been explicitly defined. 

Quality ratings aggregate the evaluation results of feature constraints on the level of interests and interest groups on an ordinal scale. Thereby, concrete fulfillment criteria for each grade on the ordinal scale can be defined. As such, they give a quick overview to external stakeholders about the degree quality-related stakeholder interests are fulfilled. 

Quality gates on the other hand summarize multiple quality ratings to determine whether a software feature or the whole software product meets specific quality criteria to proceed to the next engineering step. This decision is based on concrete pass/fail criteria which are defined using the TAICOS DSL.

\subsection{Grammar and Language Elements} 
Similarly to the previous section, when showing grammar rules for the different language elements we highlight \dslkeyword{<non-terminal symbols>} in brackets and \dslkeyword{terminal symbols} without brackets. In this context the most important language elements are \dslkeyword{<QualityRating>} and \dslkeyword{<QualityGate>}. Further, the \dslkeyword{<ProductQuality>} element allows to define a product-level quality gate and product-wide default quality ratings and gates for all software features and stakeholder interests. 

\subsubsection{Formulation and Calculation of Quality Ratings}
Quality ratings (element \dslkeyword{<QualityRating>}) can be defined for interests, interest groups, and individual software features. Depending on which of these levels a quality rating is defined, we distinguish between \textit{constraint-based} and \textit{interest-based quality ratings}. For both types of quality ratings, the calculated grade of a quality rating can be in the range of A-E, with A the best and E the worst. Listing \ref{lst:qualityratings} shows the general grammar rules for defining quality ratings.

\begin{figure}[h!]
\lstset{style=atg,basicstyle=\footnotesize\ttfamily,caption={Grammar rules for defining quality ratings.},captionpos=b,label=lst:qualityratings}
\begin{lstlisting}
<QualityRating> ::= ratingId ": { " 
  ( ConstraintBasedRating> | <InterestBasedRating> ) 
  { ( ConstraintBasedRating> | <InterestBasedRating> ) } "}" .
<ConstraintBasedRating> ::= ratingVal ":" 
  (noFulfilled "of" totalConstraints) | (percentage "of") 
  "constraints"
  [ "and" 
      ("constraint" contraintId) | 
      ("constraints" contraintId {"," contraintId })
  ] "are fulfilled." .
<InterestBasedRating> ::= "average rating of interests"
  [ interestId "," interestId { "," interestId } ] "." .
ratingVal ::= "A" | "B" | "C" | "D" .
\end{lstlisting}
\end{figure}

Constraint-based ratings (\dslkeyword{<ConstraintBasedRating>}) are calculated based on the number or percentage of fulfilled constraints for an individual stakeholder interest of a software feature. In the definition of the same quality rating one must not switch between specifying the degree of fulfillment as an absolute or relative number. Particularly, in case of only a small total number of constraints, specifying the degree of fulfillment with absolute numbers might be more suitable. On the other hand, as the number of total constraints for a particular interest might vary among different software features, defining the degree of fulfillment with relative percentages is more robust when applying the same quality rating for a particular interest to different software features.

At maximum (but not mandatory) four different grades from A-D with descending degrees of constraint fulfillment can be defined. Also, lower grades must not relax the required degree of fulfillment of higher grades. The grade E does not have to be defined explicitly, as it is automatically generated if none of the other grades are fulfilled. To achieve a particular grade, also the fulfillment of one or more specific constraints (referred to by their names) can be demanded. When calculating the grade for a quality rating, the conditions for the defined grades are evaluated in descending order until the actual degree of constraint fulfillment suffices the conditions of a grade. 
As an example, one could define a quality rating \dslkeyword{A: 4 of 4 constraints are fulfilled. B: 3 of 4 constraints and constraint technicalDebtConstraint \newline are fulfilled.} Thus, for an \dslkeyword{A}  grade, all constraints of the interest for which this quality rating was defined would need to be fulfilled. For a B grade, only 3 of the 4 constraints (one being the \dslkeyword{technicalDebtConstraint}) would need to be fulfilled. In case that this constraint is not fulfilled or less than 3 constraints are fulfilled, the resulting grade would be E, as no more further grades are defined. 

Interest-based ratings (element \dslkeyword{<InterestBasedRating>}) can be defined on two levels: If defined for an interest group (second level), the calculation of the grade takes into account constraint-based quality ratings of subordinate individual interests (first level). Contrarily, if defined on the higher level of a software feature (third level), the calculation of the grade is based on quality ratings of the feature's interest groups (second level) and individual interests (first level). In both cases, the resulting grade is calculated based on the average of either all or only selected subordinate quality ratings. Thus, for this type of quality ratings, no grades and respective constraint fulfillment can be defined, but instead a set of quality ratings as the basis for grade calculation. For calculating the grade, the ordinal grades of the subordinate quality ratings are mapped to integer values (A=1, B=2, C=3, D=4, E=5), the average calculated thereof, the resulting floating point arithmetically rounded to an integer value and mapped back to an ordinal grade. An interest-based quality rating can, e.g., be defined as \dslkeyword{average rating of interests} on a software feature with subordinate quality ratings having been graded with A, C, and D. The resulting grade for the quality rating of the software feature would then be calculated as $(A+C+D)/3 = (1+3+4)/3 = (8/3) = 2.67 \approx 3 = C$.

If no quality rating is explicitly specified, the standard grading scheme for quality ratings is applied to calculate the grades. There is an important distinction between standard and default quality ratings: \textit{Standard quality ratings} cannot be specified explicitly as they are calculated automatically in case no quality rating has been defined for the different aggregation levels by the user. In contrast, \textit{default quality ratings} can optionally be specified to set particular quality ratings as defaults for the different aggregation levels. For (standard) constraint-based quality ratings, the following percentages of constraint fulfillment are defined: A $(>=$95\%), B [80-95\%[, C [65-80\%[, D [50-65\%[, E ($<$50\%). For (standard) interest-based quality ratings, the average grade is calculated from the grades of all subordinate interest groups (for quality ratings of software features) and of all subordinate interests (for quality ratings of interest groups). 

\subsubsection{Formulation and Calculation of Quality Gates}
\label{qualitygates}
Quality gates (element \dslkeyword{<QualityGate>}) summarize multiple quality ratings and can be defined for individual software features or for the whole software product. If quality gates shall be checked they must be explicitly defined, as there is no default calculation scheme similar to quality ratings. 

The DSL provides three states for quality gates: \dslkeyword{pass}, \dslkeyword{warn}, and \dslkeyword{fail}. When defining quality gates (\dslkeyword{<GateDefinition>}), at least the criteria for the \dslkeyword{pass} state must be defined, while the definition of the \dslkeyword{warn} state is optional. The \dslkeyword{fail} state does not need to be specified as it is automatically reached if no other state applies. When evaluating quality gates, the conditions for the \dslkeyword{pass} state are checked first. If the conditions for this state are not fulfilled and yet conditions for the \textit{warn} state are defined, those are checked and upon their fulfillment this is the final state. In case this state is not defined or its conditions also are not fulfilled, the final state of the quality gate is \dslkeyword{fail}. Listing \ref{lst:qualitygates} shows the grammar rules for defining quality gates and their states using the DSL.

\begin{figure}[h!]
\lstset{style=atg,basicstyle=\footnotesize\ttfamily,caption={Grammar rules for defining quality gates.},captionpos=b,label=lst:qualitygates}
\begin{lstlisting}
<QualityGate> ::= gateId : "{ " 
  "pass:" <GateDefinition> "."
  [ "warn:" <GateDefinition> "." ] "}" .
<GateDefinition> ::= "average rating of" 
  ("interests"|"features") [referenceId { "," referenceId }] 
  "is" ratingVal 
  [ "and" 
      ( "interest" | "feature" ) 
      referenceId "is rated" ratingVal 
  ] "." .
referenceId ::= featureId | interestId .
\end{lstlisting}
\end{figure}

In general, the state of a quality gate is calculated from the average of multiple quality ratings, whereby we distinguish between \textit{interest-based gates} and \textit{feature-based gates}. Interest-based gates can be defined for individual software features and take into account quality ratings of an individual feature's subordinate interest groups and interests. Likewise, feature-based gates can be defined on the top level of the software product and take into account quality ratings of the product's software features. Similar to quality ratings, the set of quality ratings of interests and features used for calculating the average can be specified in the definition of the quality gate. Also, a particular quality rating can be demanded for an individual interest or software feature (\dslkeyword{ratingReferenceId}). To fulfill a certain state, the calculated average of the quality ratings must meet the conditions defined for this state. 

For illustration purposes, let assume that the \dslkeyword{video} feature of a fictitious software product is particularly important for the developing organization. Defining an (interest-based) quality gate as \dslkeyword{pass: average rating of interests is B.} and leaving out the optional definition of the \textit{warn} means that an average rating of the feature's quality ratings dropping below B leads to failing this quality gate. Therefore, the definition of the \textit{warn} state is intentionally left out. Further, one could define a (feature-based) quality gate for this software product as \dslkeyword{pass: average rating of features is B. warn: average rating of features is C and feature video is rated B.} Apparently, while for the \textit{warn} state the demanded average of the quality ratings is lower than for the \textit{pass} state, the \dslkeyword{video} feature nonetheless must have a quality rating better than or equal B to fulfill at least the \textit{warn} state.

\subsubsection{Defining Default Quality Ratings and Gates}
The \dslkeyword{<ProductQuality>} element allows to define default quality ratings, default quality gates, and the top-level (feature-based) quality gate for the software product. Defaults prevent redundant specifications and can be defined for 
\begin{itemize}
\item quality gates for software features (interest-based gates)
\item quality ratings for software features (interest-based ratings)
\item quality ratings for interests (constraint- \& interest-based ratings)
\end{itemize}
Depending on whether default quality ratings and gates are defined for the software product, individual features or interests (and groups), different types of quality ratings and gates are applicable. As an example, for software features quality ratings and gates require the subordinate interest-based counterparts for their calculation. On the level of interests, constraint- and interest-based quality ratings can be defined as defaults. The former to calculate quality ratings through constraint fulfillment on interest-level (\dslkeyword{constRatingId}), the latter from quality ratings of multiple subordinate interests of an interest group (\dslkeyword{intrstRatingId}). As the total number of constraints might vary among different interests, for default constraint-based quality ratings the required level of constraint fulfillment shall be described in percentage.

The defaults are applied to all features and interests by the TAICOS DSL parser when instantiating the TAICOS model. However, these defaults can be overridden individually by software features or interests.

%% file: example.tex
\label{example}
To make the language elements of the TAICOS DSL more tangible, in this section we present an illustrative example based on sample quality requirements for the secure mobile messaging app \textit{Telegram}\footnote{https://telegram.org/}. We intentionally picked this app due to its popularity and its open source policy and selected four different software features with different code sizes. While the \textit{video} and \textit{camera} feature might be self-explanatorily, the \textit{voip} feature allows to make voice calls using the cell phone's mobile data connection and the \textit{secretmedia} feature adds additional security measures for chats and exchanging media files.
\begin{figure*}[h!]
\centering
\includegraphics[width=\textwidth]{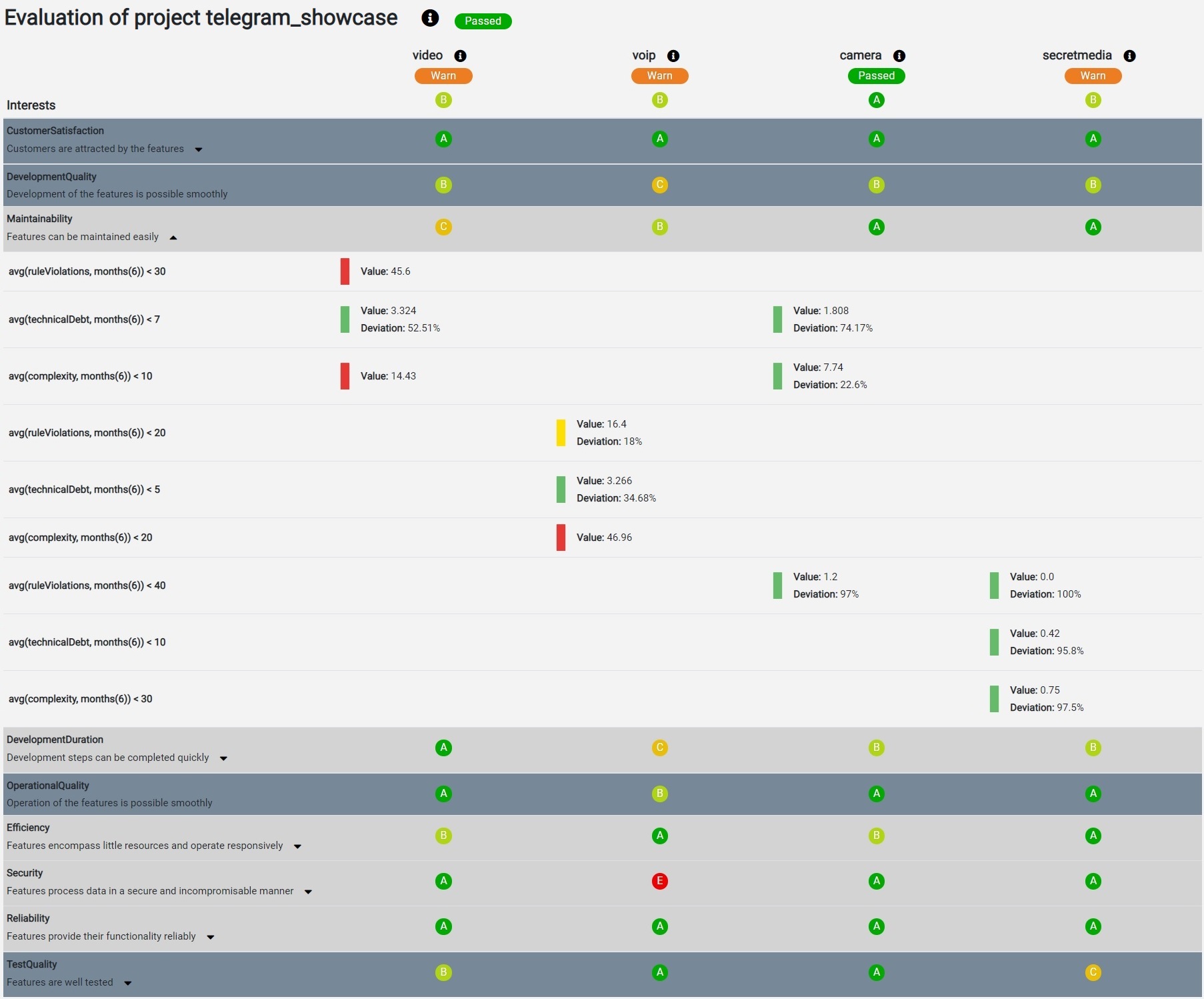}\caption{TAICOS web application showing aggregated constraint evaluation results.\label{fig:taicos_webapp}}
\end{figure*}
 We formulated seven stakeholder interests and two interests groups which are generally representative for such an application. Figure \ref{fig:taicos_webapp} shows a screenshot of the TAICOS web application with the four features horizontally at the top and the interests (and groups) with the corresponding constraints for all features on the left side. The web application displays interests with a downward arrow next to their description to collapse their corresponding constraints. Thereby, the background color indicates whether the respective interest is a self-contained interest (blue) or is a subordinate interest (gray) of an interest group. Interest groups are displayed without a downward arrow with blue background. For instance, the interests \textit{Maintainability} and \textit{DevelopmentDuration} belong to the interest group \textit{DevelopmentQuality}, whereas \textit{CustomerSatisfaction} is a self-contained interest.

As each of the four features also needs to individually fulfill the quality-related objective of each interest, we specified respective feature constraints. For this demonstration, we captured metrics from 21 different versions (4.9.0 - 6.3.0) of the \textit{Telegram} app, which were released during July 2018 - July 2020. We operationalized these metrics from, e.g., commit metadata of the app's official \textit{GitHub} repository, customer ratings from the \textit{Google Play Store}, static code analysis results, and operational data from instrumenting the source code. 

In the context of this illustrative example, the presented constraints primarily intend to demonstrate the capabilities of the DSL and not to thoroughly assess the overall product quality of the \textit{Telegram} app. Finally, we also defined individual quality ratings for five selected interests and quality gates on feature- and product-level. The cells of the matrix show the fulfillment of constraints defined for a particular feature. For the sake of brevity, we only show the feature constraints for the \textit{Maintainability} interest. As an example, the constraint \dslkeyword{avg(technicalDebt), months(6) < 7} is defined with this threshold (\dslkeyword{<7}) only for the \textit{video} and \textit{camera} feature, whereas it is stricter formulated (\dslkeyword{<5}) for the \textit{voip} feature and more relaxed (\dslkeyword{<10}) for the \textit{secretmedia} feature. For this illustrative example, we defined a \dslkeyword{warningThreshold} of 20\% (cf. Section \ref{warningThreshold}). Colored rectangles next to the constraints indicate their fulfillment. Thereby, violated constraints are colored in red, constraints fulfilled above the warning threshold in green and constraints yet fulfilled but below the warning threshold in yellow. For instance, the constraint of the \textit{Maintainability} interest regarding rule violations of the \textit{voip} feature yet is fulfilled, but with 18\% below the warning threshold. Colored circles show the grades of the calculated quality ratings for interest groups, interests, and software features. The state of a quality gate is shown in rounded rectangles above a feature's quality rating. In case of the software product it is shown next to the project name in the left upper part.
\vspace{-2pt}
\begin{figure}[h!]
\lstset{style=atg,basicstyle=\footnotesize\ttfamily,caption={Quality ratings and gates defined for the Telegram showcase.},captionpos=b,label=lst:example}
\begin{lstlisting}
@quality-gates: {
   product_quality_gate: {
      pass: average rating of features is B.
      warn: average rating of features is C and 
            feature video is rated B.
   }
   feature_quality_gate: {
      pass: average rating of interests is A.
      warn: average rating of interests is B. } 
}
@quality-ratings: {
   maintainability_rating: {
      A: 3 of 3 constraints are fulfilled.   
      B: 2 of 3 constraints and 
         constraint technicalDebtConstraint are fulfilled.
      C: 1 of 3 constraints are fulfilled. } 
}
\end{lstlisting}
\vspace{-2pt}
\end{figure}

Listing \ref{lst:example} shows the formulated quality rating for the \textit{Maintainability} interest, the quality gate \dslkeyword{feature\_quality\_gate} as default quality gate for all software features, and the quality gate \dslkeyword{product\_quality\_gate} for the overall software product. In the following, we only focus on the calculation of quality ratings and gates for the \textit{video} feature and the software product. Figure \ref{fig:taicos_webapp} shows that 1 of the 3 constraints of the \textit{Maintainability} interest are fulfilled for this feature, the one fulfilled constraint being the \dslkeyword{technicalDebtConstraint}. Based on the definition of the quality rating \dslkeyword{maintainability\_rating} for this interest, this results in a C grade. As can be derived from its background color, this interest is a subordinate interest of the interest group \textit{DevelopmentQuality}. As we defined no specific quality rating for this interest group, it is calculated as the average of (sub)interests' quality ratings by default, i.e., as $(C + A)/2 = (3 + 1)/2 = 4/2 = B$. Similary, the quality rating of the \textit{video} feature is calculated from the average of its directly subordinate interest groups and interests. Concretely, this comprises the quality ratings of the self-contained interests \textit{CustomerSatisfaction} and \textit{TestQuality} and of the interest groups \textit{DevelopmentQuality} and \textit{OperationalQuality}, and results in $(A + B + B + A)/4 = (1 + 2 + 2 + 1)/4 = 6/4 = 1.5 \approx 2 = B$. 

The quality gate \dslkeyword{feature\_quality\_gate} is defined as default quality gate for all four features. It determines that the \dslkeyword{pass} state is only achieved if a feature has an average rating of A, whereas upon an average of B it reaches the \dslkeyword{warn} state. These conditions are quite strict, as we want to ensure that the quality of the software features is uniformly at the highest level, and deviations from this goal therefore are immediately reflected in the quality gates. Hence, only the gate for the \textit{camera} feature fulfills the condition for the \dslkeyword{pass} state, the other three features' gates are in \dslkeyword{warn} state. On the level of the software product, however, the conditions are more relaxed. The \dslkeyword{product\_quality\_gate} nonetheless demands at least a B rating for the \textit{video} feature, due to its anticipated relevance in this example. As a result, the quality gate for the overall software product results in a \dslkeyword{pass} state. The different strict criteria of these quality gates also show how contrasting levels of criticality can be enforced on software features and the overall software product, using the DSL.

%% file: discussion.tex
\label{discussion}
For the practical applicability of a DSL for expressing feature-level quality requirements and aggregating the evaluation results, two key factors need to be addressed. First, context-dependent usage patterns of the language might impose specific performance requirements towards software tools implementing the language. These usage patterns are associated, e.g., with different types of quality requirements and can result in some language elements being more frequently used than others. Second, the DSL must be suitable, expressive, and complete from the perspective of practitioners to address quality requirements. Missing important basically expected capabilities towards the language from practitioners, inevitably hinders its practical adoption or leads to complicated and unintended use of language elements.

We analyzed the performance of the different language elements based on 12 releases (over 1 year) of the widely adopted \textit{Elasticsearch}\footnote{https://www.elastic.co/elasticsearch/} search engine in a lab experiment \cite{haindl_operational_2020}. Thereby we used quality measures from static code analysis which are typical for large-scale software products. While performance characteristics in themselves do not depend on the DSL but on the language's software implementation, each time series operation inherently requires particular mathematical operations. The validation showed that even if the time series operations take into account measures of over one year and of more than 800 classes, they finish within 1 millisecond with the \dslkeyword{gradient} operation taking the longest. However, the most time-consuming part during constraint evaluation attributes to the retrieval of the measures, which basically depends on the concrete software implementation of the language. In \cite{haindl_specifying_2020} we presented our case study with practitioners who applied the language to formulate feature constraints for maintainability requirements. In this context, we identified the \dslkeyword{measure} data type as being most frequently used and the \dslkeyword{min}, \dslkeyword{max}, \dslkeyword{median}, and \dslkeyword{avg} time series operations used far more often than the \dslkeyword{gradient} and \dslkeyword{benchmark} operations. 

In contrast to evaluating feature constraints, far less data are required for aggregating evaluation results either as quality ratings or gates. This is due to both quality ratings and gates relying on already calculated evaluation results to thereof calculate the average. Therefore, we can assume that these language elements will not have a significant impact on the overall time to evaluate feature constraints and aggregate their evaluation results.

Specially to facilitate the definition and interpretation of quality ratings in practice, we have followed the widely used 5-level ordinal scale which is also used in the \textit{Sonarqube}\footnote{https://www.sonarqube.org/} suite, for example. We have also taken deliberate care to avoid complexity in the definition of quality ratings and gates. This means that both types of aggregations shall be applicable by people who are less familiar with technical details (product managers, project managers), in contrast to people who define the feature constraints (software engineers, architects). Nonetheless, as also expressed by the participants of our case study \cite{haindl_specifying_2020}, it is important to be able to access evaluation results of individual constraints, interests, or features in the definition of quality ratings and gates. Also, the translation of quality requirements to each individual feature should be done exclusively through feature constraints, since at the higher levels (i.e. interests groups and features) only subordinate evaluation results are aggregated.

%% file: conclusion.tex
\label{conclusion}
This paper presented an operational DSL for expressing feature-level quality requirements and the aggregation of evaluation results for individual software features and the overall software product. In this context, \textit{operational} refers to defining these quality requirements in a manner that allows their automated evaluation of fulfillment and the aggregation of these results in DevOps. Our DSL allows to define these quality requirements individually for each software feature as feature constraints, which are based on quantitative measures. The main idea behind our approach is that quality requirements in general have different criticality for software features and thus, must also be tailored and individually specified with appropriate thresholds for each software feature. Using our DSL, not only the acquisition of these measure from external sources in DevOps can be defined, but also their filtering, preprocessing, and evaluation using different time filters, time series operations, and comparison operators. 

Also, we introduced two novel language elements to define the aggregation of evaluation results. When conceptualizing these two language elements and their typical usage scenarios, we incorporated our findings from a previous interview study in which the participants addressed the need for these language capabilities. For specifying results aggregation, we distinguish between quality ratings and quality gates. The former summarize the evaluation results of each software feature based on the average of these evaluation results on a five-point ordinal grading scheme from A-E. Likewise, the latter use these quality ratings to reflect the fitness of individual features or the overall software product through pass, fail, or warning states. Each degree of fulfillment of quality ratings and gates can be described using the DSL. Quality ratings can also demand the fulfillment of individual constraints or particular grades of subordinate quality ratings. Similarly, quality gates can also demand that a software feature reaches a certain grade of quality rating.

Future work needs to concentrate on empirical validations regarding the suitability of the novel two language elements. This particularly pertains to the expressiveness and completeness of the language to formulate conditions for the individual grades of quality ratings and to define the different states of quality gates. Also, methodological and tool support shall be improved to better support engineering teams in defining feature-level quality constraints, quality ratings and gates. Methodological support shall specially cover the elicitation of quality requirements and their individual feature-dependent relevance from the different stakeholders in DevOps, based on the domain of the software product.